\documentclass[letterpaper, 10 pt, conference]{ieeeconf}  

\IEEEoverridecommandlockouts                              

\usepackage[english]{babel}
\usepackage[utf8]{inputenc}
\usepackage[cmex10]{amsmath}
\usepackage{amssymb}

\usepackage{float}
\usepackage[pdftex]{graphicx}
\usepackage{amsfonts}
\usepackage{mathtools}
\usepackage{xcolor}
\usepackage{color}

{}
{}
\newtheorem{proposition}{Proposition}{}
\newtheorem{theorem}{Theorem}{}
\newtheorem{remark}{Remark}{}
\newtheorem{lemma}{Lemma}{}

\title{\LARGE \bf
A Comparison of Stealthy Sensor Attacks on Control Systems
}

\author{Navid Hashemi$^{1}$, Carlos Murguia$^{2}$, and Justin Ruths$^{1}$
\thanks{*This work was partially supported by the National Research Foundation (NRF), Prime Minister's Office, Singapore, under its National Cybersecurity R\&D Programme (Award No. NRF2014NCR-NCR001-40) and administered by the National Cybersecurity R\&D Directorate.}
\thanks{$^{1}$These authors are with the Departments of Mechanical and Systems Engineering at the University of Texas at Dallas, Richardson, Texas, USA 
        {\tt\small Navid.Hashemi, jruths @utdallas.edu}}%
\thanks{$^{2}$C. Murguia is with the iTrust Centre at the Singapore University of Technology and Design, Singapore
        {\tt\small murguia\_rendon@sutd.edu.sg}}%
}

\begin{document}

\maketitle
\thispagestyle{empty}
\pagestyle{empty}

\begin{abstract}
As more attention is paid to security in the context of control systems and as attacks occur to real control systems throughout the world, it has become clear that some of the most nefarious attacks are those that evade detection. The term \textit{stealthy} has come to encompass a variety of techniques that attackers can employ to avoid detection. Here we show how the states of the system (in particular, the reachable set corresponding to the attack) can be manipulated under two important types of stealthy attacks. We employ the chi-squared fault detection method and demonstrate how this imposes a constraint on the attack sequence either to generate no alarms (zero-alarm attack) or to generate alarms at a rate indistinguishable from normal operation (hidden attack).

\end{abstract}


\section{INTRODUCTION}
For many decades, Control Theory operated in a challenging but happy place in which problems pitted designers against the world, a haphazard place of disturbances and uncertainty. The past decade has seen the rise in concern over attacks on control systems, which necessarily requires us to shift our focus to a problem of designer against attacker, a strategic and knowledgeable entity that seeks to exploit the weaknesses of our systems and control frameworks. 

Control systems have become an attractive target to attackers due to accessibility, impact, and obfuscation. Large-scale control systems such as process control plants are increasingly moving toward Ethernet-like technology to communicate data throughout the system. This new architecture provides new capabilities but also opens systems up to the same types of cyber attacks that banking and database companies endure. These systems also represent major industry or municipal infrastructure, which means damaging them makes large impact. Finally, these systems are large and complex enough - and often not monitored well enough - for attackers to manipulate the system without being detected.

The literature of attack detection has concerned itself with designing methods to effectively monitor systems and detect anomalies \cite{Cardenas}\nocite{Pasqualetti_1}\nocite{Mo_1}\nocite{Kwon}\nocite{Pappas}-\cite{Gupta2}. The origin of many of these methods arise from fault detection, but have been retooled to consider antagonistic and strategic ``faults''. A key component of this body of work is to understand the limits of these detectors, and identifying attacks that are stealthy to these methods is a critical way to benchmark detector performance. The term \textit{stealthy} has taken on several meanings in the literature. It has been used to address attacks that do not induce the detectors to raise alarms; we rename these \textit{zero-alarm attacks} to be more precise \cite{Cardenas,Carlos_Justin1,Carlos_Justin2}. It has also referred to an attack that changes the alarm rate of the detector by only a small amount; we call these \textit{perturbation attacks} \cite{Mo_3,Carlos_Justin3}. We define a \textit{hidden attack} which exactly mimics the alarm rate of the detector. Stealthy is also used to describe attacks that effect the uncontrollable and unobservable modes of the system and, therefore, do not propagate to any measurement or estimated state of the system \cite{Pasqualetti_1}. Replay attacks also fall into the category of stealthy attacks as they replay past (recorded) data back to the monitoring equipment \cite{Mo_2}.

The attacks on unobservable/uncontrollable modes and replay attacks completely circumvent the detectors, which is interesting and relevant to the broader context of security, but requires a countering strategy that goes beyond detectors. The perturbation attack has a relatively small effect on the system compared with zero-alarm and hidden attacks, thus we also omit it from this study. 

We use this manuscript to present a distribution-based perspective on attack detection. While this in and of itself is not novel, this way of looking at and describing attacks has yet to be captured clearly in the literature. As part of this we present a equitable comparison between the impact of zero-alarm and hidden attacks. We use the set of states reachable by the system when driven by the attacker input as a metric for this comparison. To achieve this, we present several novel results on techniques to formulate and algorithms to find ellipsoidal outer bounds on the reachable sets of the system corresponding to attacks.


\section{BACKGROUND}
In this work, we study stochastic discrete-time linear time-invariant (LTI) systems
\begin{equation} \label{eq:dLTIsystem}
\left\{\begin{aligned}
	x_{k+1} &= Fx_k + Gu_{k} + v_{k},\\
	y_k &= Cx_k + \eta_k, \\
\end{aligned}\right.
\end{equation}
in which the state $x_k\in\mathbb{R}^n$, $k\in\mathbb{N}$, evolves due to the state update provided by the state matrix $F\in\mathbb{R}^{n\times n}$, the control input $u_k\in\mathbb{R}^m$ filtered by the input matrix $G\in\mathbb{R}^{n\times m}$, and the i.i.d. zero-mean Gaussian system noise $v_k$ with covariance matrix $R_1$. The output $y_k\in\mathbb{R}^p$ aggregates a linear combination, given by the observation matrix $C\in\mathbb{R}^{p\times n}$, of the states and zero-mean Gaussian measurement noise with covariance matrix $R_2$.  We assume that the pair $(F,C)$ is detectable and $(F,G)$ is stabilizable.

In this work, we consider the scenario that the actual measurement $y_k$ can be corrupted by an additive attack, $\delta_k \in \mathbb{R}^p$. At some point in the process of measuring and transmitting the output to the controller the attacked output becomes
\begin{equation}
	\bar{y}_k = y_k + \delta_{k} =  Cx_{k} + \eta_{k} + \delta_k.
\end{equation}
If the attacker has access to the measurements, then it is possible for the attack $\delta_k$ to cancel some or all of the original measurement $y_k$ - so an additive attack can achieve arbitrary control over the ``effective'' output of the system.

As our approach leverages a fault-detection approach, we require an estimator of some type to produce a prediction of the system behavior. In this work we use the steady state Kalman filter
\begin{equation}
	\hat{x}_{k+1} = F\hat{x}_k + Gu_k + L(\bar{y}_k - C\hat{x}_k),
\end{equation}
where $\hat{x}_k \in \mathbb{R}^n$ is the estimated state. The observer gain $L$ is designed to minimize the steady state covariance matrix $P:= \lim_{k \rightarrow \infty}P_k:= E[e_ke_k^T]$ in the absence of attacks, where $e_k:= x_k - \hat{x}_k$ denotes the estimation error. Existence of $P$ is guaranteed since the pair $(F,C)$ is assumed to be detectable \cite{Astrom}. Next, we define the residual sequence $r_k$
\begin{equation}
	r_k := \bar{y}_k - C\hat{x}_k,
\end{equation}
the difference between what we actually receive ($\bar{y}_k$) and expect to receive ($C\hat{x}_k$), which evolves according to
\begin{equation} \label{eq:esterror}
\left\{ \begin{aligned}
	e_{k+1} &= \big( F - LC \big) e_k - L\eta_k + v_k - L \delta_k, \\
	r_k &= Ce_k + \eta_k + \delta_k.
\end{aligned}\right.
\end{equation}
In the absence of attacks (i.e., $\delta_k = 0$), it is straightforward to show that the $r_k$ random variable falls according to a zero mean Gaussian distribution with covariance \cite{Carlos_Justin2}
\begin{equation}
	\Sigma = E[r_kr_k^T] = CPC^T+R_2.
\end{equation}

In this work, we consider only one detector, the popular chi-squared detector. Although other alternatives exist, the chi-squared is easily the dominant choice for most research and it also provides a transparent choice to highlight the key messages we wish to communicate in this work. Similar analysis can be done with these other detector choices using attacks derived in our other work \cite{Carlos_Justin1,Carlos_Justin2,RuthsACC_Windowed}. In the case of the chi-squared detector, a quadratic distance measure $z_k$ is created to be sensitive to changes in the variance of the distribution as well as the expected value, 
\begin{equation}
	z_k = r_k^T\Sigma^{-1}r_k.
\end{equation}
Since $r_k \sim \mathcal N(0,\Sigma)$, the $z_k$ random variable, as the sum of the squares of normally distributed random variables, falls according to the chi-square distribution. Since $r_k\in\mathbb{R}^p$, this chi-squared distribution has $p$ degrees of freedom. The chi-squared detector is summarized as follows: for given a threshold $\alpha \in \mathbb{R}_{>0}$ and the distance measure \linebreak $z_k = r_k^T\Sigma^{-1}r_k$
\begin{equation} 
\left\{\begin{aligned}
	z_k \leq \alpha &\quad\longrightarrow\quad \text{no alarm}, \\
	z_k > \alpha &\quad\longrightarrow\quad \text{alarm: }k^* = k,
\end{aligned}\right.
\end{equation}
alarm time(s) $k^*$ are produced. The $\Sigma^{-1}$ factor in the definition of $z_k$ rescales the distribution ($E[z_k]=p$, $E[z_kz_k^T]=2p$) so that the threshold $\alpha$ can be designed independent of the specific statistics of the noises $v_k$ and $\eta_k$; instead, it can be selected simply based on the number of sensors (i.e., the dimension of the output, $p$).

It is important to note that because of the infinite support of noises $v_k$ and $\eta_k$, the distance measure $z_k$, distributed according to a chi-squared distribution, also has infinite support. Therefore, even in the absence of attacks, we expect that the detector will generate alarms because some values drawn from the distance measure distribution will exceed the threshold $\alpha$. Such alarms in the absence of an attack are called \textit{false alarms}. Because we can characterize the chi-squared distribution analytically, we have an exact relation between the choice of the threshold $\alpha$ and the expected rate of false alarms $\mathcal{A}$ generated by the chi-squared detector.

\begin{lemma} \cite{Carlos_Justin2}. \label{lem:chisquared_tuning}
Assume that there are no attacks to the system and consider chi-squared detector, with threshold $\alpha \in \mathbb{R}_{>0}$, $r_k \sim N(0,\Sigma)$. Let $\alpha = \alpha^{*}:=2P^{-1}(1-\mathcal A^{*}, \frac{p}{2})$, where $P^{-1}(\cdot,\cdot)$ denotes the inverse regularized lower incomplete gamma function, then ${\mathcal A}= {\mathcal A}^{*}$. 
\end{lemma} 

\subsection{Undetected Attacks}
Some of the most insidious attacks on industrial control systems feature attack strategies that manipulate the system while all the time staying undetected. The effect of the attack can aggregate during this ``stealthy'' execution of the attack and the damage caused by the attack can spread. If obvious attacks were used, single components might be damaged, but it would give operators the opportunity to react in time to prevent further damage. When attacks are undetected, single damaged components might lead to other components being damaged without operators realizing the changes to the system. Past attacks on industrial control systems seem to favor these undetected attacks, such as the famous Stuxnet worm incident \cite{langner2011stuxnet}.

In many industrial settings fault detection is accomplished simply by assigning a collection of static rules (e.g., if a pressure in a vessel exceeds a given value). These offer little-to-no protection against stealthy adversarial attacks as the attack can deviate the actual system state while reporting a state that is within normal operating conditions. When detectors are implemented in control systems, these detectors limit what the attacker is able to accomplish if he/she seeks to remain undetected. We advance two notions of undetected attacks (we phrase these with respect to the chi-squared detector, however, the concept of these attack classes generalize to other detectors). These attack models require strong attacker knowledge and access, namely we assume that the attacker has perfect knowledge of the system dynamics, the Kalman filter, control inputs, measurements, and chi-squared procedure. In addition, the attacker has read and write access to all the sensors at each time step. The goal of these stealthy attacks is to construct a worst case scenario. In the same spirit of designing buildings for a 1000-year earthquake, we aim to design the control infrastructure against a strong opponent. If designers and operators are comfortable with the security performance given this kind of strong attacker, they will also accept the performance for less powerful attackers.

\begin{enumerate}
\item \textit{Zero-alarm attacks} generate attack sequences that maintain the distance measure at or below the threshold, i.e., $z_k \leq \alpha$. These attacks generate no alarms during the attack. To satisfy this condition we define the attack as
\begin{equation}\label{eq:deltabar}
	\delta_k = - Ce_k-\eta_k+\Sigma^{\frac{1}{2}}\bar{\delta}_k,
\end{equation}
where $\bar\delta_k\in\mathbb{R}^p$ is any vector such that $\bar\delta_k^T\bar\delta_k \leq \alpha$ and $\Sigma^{\frac{1}{2}}$ is the symmetric square root of $\Sigma$ (recall the attacker has read access to the sensor, $y_k$, and knowledge of the estimator, $\hat{x}_k$).  This attack sequence leads the distance measure to become
\begin{align}
	z_k &= r_k^T\Sigma^{-1}r_k \nonumber\\
    &= (Ce_k+\eta_k+\delta_k)^T\Sigma^{-1}(Ce_k+\eta_k+\delta_k) \nonumber\\
    &= (\Sigma^{\frac{1}{2}}\bar\delta_k)^T\Sigma^{-1}(\Sigma^{\frac{1}{2}}\bar\delta_k) \leq \alpha. \label{eq:zeroalarmdef}
\end{align}
Since $z_k\leq\alpha$, no alarms are raised. A schematic of a zero-alarm attack is shown in Fig. \ref{fig:attack_distributions}. Although generating no alarms seems like a successful strategy to avoid detection, it is important to remember that in the attack-free case alarms are raised due to the infinite support of the distance measure distribution. Thus, before the attack, alarms are raised at a rate $\mathcal{A}>0$ and after the attack the alarm rate becomes zero, $\mathcal{A}=0$. While the detector does not monitor changes in the false alarm rate, it is possible that an operator might notice this discrepancy. This leads us to develop a second class of undetectable attacks. We are also motivated to develop the following attacks because they exploit the stochasticity to inject larger, more potent attacks. 

\begin{figure}[t]
\label{nchi}
\centering
\label{fig:attack_distributions}
\includegraphics[width=0.9\linewidth]{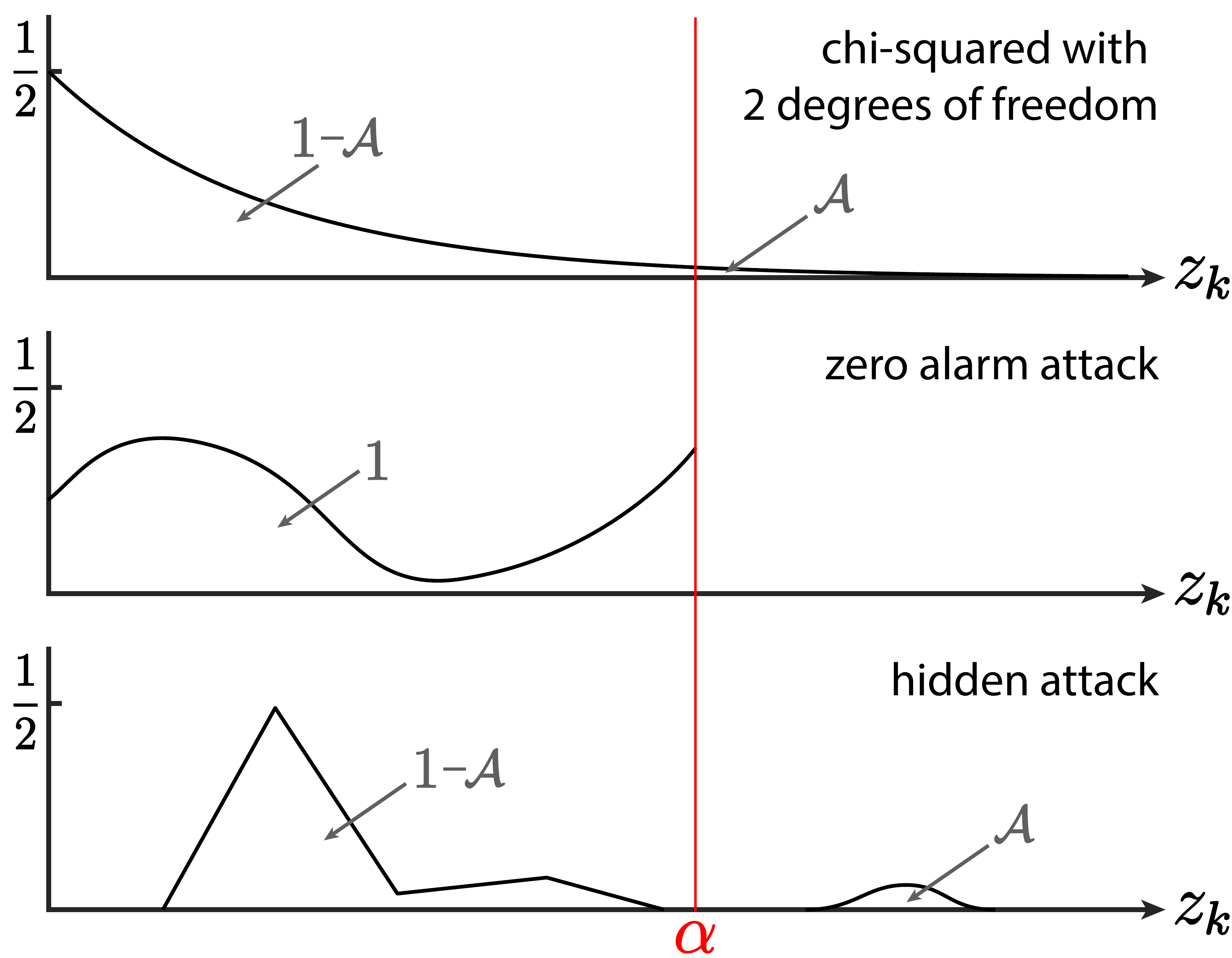}
\centering
\caption{The original (attack-free) $z_k$ distribution (top) is chi-squared with $p$ degrees of freedom (this paper uses examples in which $p=2$). The threshold $\alpha$ is selected to satisfy a false alarm rate of $\mathcal{A}$, implying that in the attack-free distribution, the area under the distribution curve that falls beyond $\alpha$ is $\mathcal{A}$. In zero-alarm attacks (middle), $\bar\delta_k$ is selected such that $z_k$ is no larger than $\alpha$, implying that no alarms are raised under zero-alarm attacks. In hidden attacks (bottom), $\bar\delta_k$ is designed so that the fraction of the distribution that falls beyond $\alpha$ matches that of the attack-free distribution, which means that the alarm rate under the hidden attack is equal to the false alarm rate. The definition of the zero alarm and hidden attacks do not stipulate the shape of the density functions above and below $\alpha$, although the allocation of mass in the density function greatly influences the effect of the attack on the reachable states. }
\end{figure}

\item \textit{Hidden attacks} generate attack sequences that raise alarms at the same rate as the false alarm rate of the detector (i.e., alarms are raised at the same rate during the attack as are false alarms in the attack-free case). In hidden attacks, the attack sequence $\bar\delta_k$ in \eqref{eq:deltabar} is a random variable designed such that
\begin{equation} \label{eq:hiddendef}
	\text{Pr}(z_k > \alpha) = \text{Pr}(\bar\delta_k^T\bar\delta_k > \alpha) = \mathcal{A}.
\end{equation}
In other words, on average out of $N$ time steps: $(1-\mathcal{A})N$ time steps $\bar\delta_k^T\bar\delta_k \leq \alpha$ and the remaining $\mathcal{A}N$ time steps $\bar\delta_k^T\bar\delta_k > \alpha$. The chi-squared detector tuned to a false alarm rate $\mathcal{A}$ effectively splits the $z_k$ distribution into a part $z_k\leq\alpha$ and a part $z_k>\alpha$, where $\alpha$ is selected using Lemma \ref{lem:chisquared_tuning}. Hidden attacks ensure that the proportion of $z_k$ values larger than $\alpha$ observed by the detector during the attack match the proportion expected in the attack-free case \cite{RuthsACC_Hidden}. A schematic of a hidden attack is shown in Fig. \ref{fig:attack_distributions}.
\end{enumerate}

\subsection{Feedback}

In order for the attack to propagate from the estimation error to the state, we need to incorporate a model of feedback in the control system. In this paper we assume static estimator feedback $u_k=K\hat{x}$. With this feedback the closed-loop system becomes
\begin{equation}
\left\{\begin{aligned} \label{eq:state_error}
	x_{k+1} &= (F+GK)x_k+G K e_k+v_k,\\
	e_{k+1} &= (F-LC)e_k-L\delta_k-L\eta_k+v_k.\\
\end{aligned}\right.
\end{equation}
The estimation error updates according to, without attacks,
\begin{equation} \label{eq:error_noattack}
	e_{k+1} = (F-LC)e_k-L\eta_k+v_k,
\end{equation}
and with attacks of the form in \eqref{eq:deltabar},
\begin{equation} \label{eq:error_attack}
	e_{k+1} = Fe_k-L\Sigma^{\frac{1}{2}}\bar\delta_k+v_k.
\end{equation}

\begin{remark} Note that if the spectral radius $\rho[F] > 1$, then $\|E[e_k]\|$ (and also $\|E[x_k]\|$ due to the interconnection) diverges to infinity as $k$ grows for any non-stabilizing $k$. That is, attacks of the form \eqref{eq:deltabar} may destabilize the system if $\rho[F] > 1$. If $\rho[F]\leq 1$, then $\|E[e_k]\|$ may or may not diverge to infinity depending on algebraic and geometric multiplicities of the eigenvalues on the unit circle. Thus we consider open-loop stable system matrices, $\rho[F]<1$.
\end{remark}

\section{REACHABLE SET BOUNDS}
In order to compare the effects of these different stealthy attacks, we require a metric to quantify the impact of each attack. A popular choice to quantify system impact due to a disturbance is the set of states reachable by the action of the disturbance. Here, we show two techniques to derive outer ellipsoidal bounds on the reachable states. The first, based on Linear Matrix Inequalities (LMIs), constructs a convex optimization problem, the solution of which is the ellipsoid that bounds the states driven by attacks.  This approach provides more conservative estimates of the reachable set, but allows for the opportunity to simultaneously design system components, such as estimator and controller gain matrices, to reduce the size of the reachable set (see, e.g., \cite{Carlos_Justin3,RuthsACC_Hidden}). The second approach provides extremely tight bounds for the reachable set through the use of geometric ellipsoidal methods.

\begin{figure}[t]
\centering
\includegraphics[width=0.8\linewidth]{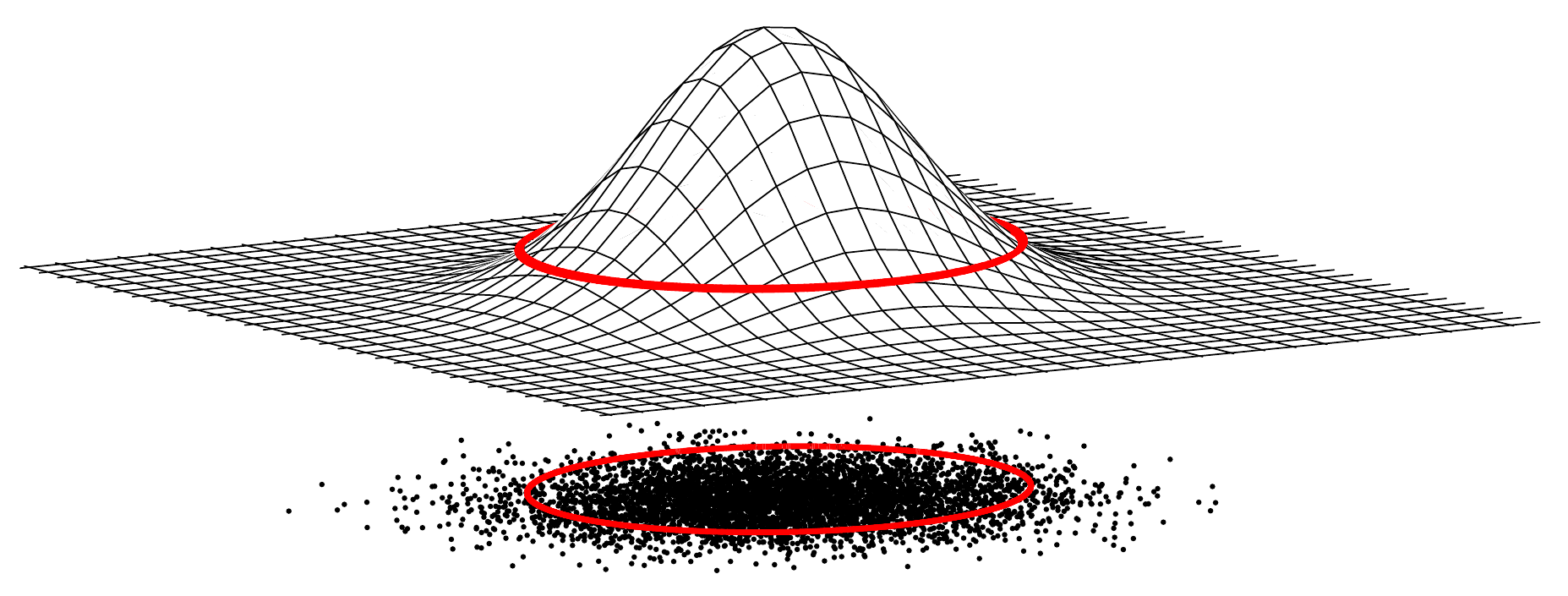}
\caption{The $\bar{p}$-probable ellipsoid captures a level set of the distribution of reachable states corresponding to when the system is driven by a truncated distance measure, such that $z_k\leq\bar{z}$, where $\text{Pr}(z_k\leq \bar{z})=\bar{p}$.} \label{fig:pprobable}
\end{figure}

The different definitions of the zero alarm attack and the hidden attack naturally give rise to different reachable sets. The challenge of the hidden attack definition is that there are no constraints on the location of the $\mathcal{A}$ mass that falls beyond $\alpha$. This means that $\mathcal{A}$ of the probability density function of the distance function can be made arbitrarily large, which in turn makes the reachable sets driven by hidden attacks arbitrarily large. Therefore, it is not meaningful to define the outer bound of the reachable set corresponding to hidden attacks - it would simply be the entire state space. Instead, we introduce the notion of a $\bar{p}$-probable ellipsoid which encompasses the reachable set when the $z_k$ distribution is truncated at $\bar{z}$, where $\text{Pr}(z_k\leq \bar{z})=\bar{p}$. This ellipsoid can be interpreted graphically as a level set of the distribution function of the reachable states (see Fig. \ref{fig:pprobable}). It is worth noting that the probability $\bar{p}$ corresponds to the truncation of the $z_k$ distribution and does not specify the probability that a point in the true reachable state set is in the $\bar{p}$-probable ellipsoid. While the later probability is closer to what we want to know, this would require knowing the complete distribution of reachable states which is what we aim to find in the first place. Notwithstanding, there is a one-to-one mapping from $\bar{p}$ to level sets of the reachable set distribution, so increasing (resp., decreasing) $\bar{p}$ necessarily expands (resp., shrinks) the ellipsoid level set, so this is not much of a restriction.

While we consider more general choices of $\bar{p}$ in other work (see \cite{RuthsACC_Hidden}), here we focus on the most immediate choice $\bar{p}=1-\mathcal{A}$ and so $\bar{z}=\alpha$. Recall that the hidden attack (see Fig. \ref{fig:attack_distributions} and \eqref{eq:hiddendef}) only requires the attacker to satisfy one statistic of the attack $\bar\delta_k$, namely $\text{Pr}(z_k=\bar\delta^T_k\bar\delta_k\leq\alpha)=1-\mathcal{A}$. For a general hidden attack, we have no further information about the distribution of $z_k$ (recall the shape of the distribution, beyond this one constraint, is completely free for the attacker to choose, see Fig. \ref{fig:attack_distributions}). Thus the only choice of $\bar{p}$ that can be evaluated for a general hidden attack is $\bar{p}=1-\mathcal{A}$. This also simplifies the comparison of hidden attacks with zero alarm attacks. Selecting $\bar{z}=\alpha$ as the truncation point of the $z_k$ distribution implies then that we truncate the attacks such that $z_k=\bar\delta^T_k\bar\delta_k\leq\alpha$. This truncation to quantify the $\bar{p}$-probable ellipsoidal bound for the reachable set due to hidden attacks now imposes the same constraint that exists in the case of zero alarm attacks, see \eqref{eq:zeroalarmdef}.

When we look at the complete reachable state of the system, we can decompose the contributions due to system noise and due to attack separately. Using the superposition principle of linear systems, the estimation error $e_k$ can be written as $e_k = e_{k}^v + e_{k}^\delta$,  where $e_{k}^v$ denotes the part of $e_k$ driven by noise and $e_{k}^\delta$ is the part driven by attacks. We can now write the estimation error dynamics in \eqref{eq:error_attack} as follows, where we assume the attack starts at $k=k^*$,
\begin{align}
&e_{k+1}^v = Fe_{k}^v + v_k, \qquad\qquad e_{k^*}^v=e_{k^*} \label{eq:estimation_noise}\\
&e_{k+1}^\delta = Fe_{k}^\delta -  L \Sigma^{\frac{1}{2}} \bar{\delta}_k,\qquad e_{k^*}^\delta=\mathbf{0}.\label{eq:estimation_attack}
\end{align}
Similarly, the state of the system $x_k$ can be written as $x_k = x_{k}^v + x_{k}^\delta$,  where $x_{k}^v$ denotes the part of $x_k$ driven by noise and $x_{k}^\delta$ is the part driven by attacks. Using this new notation, we can write the system dynamics in \eqref{eq:state_error} as follows,
\begin{align}
&{x}_{k+1}^v = (F + GK)x_{k}^v - GK e_{k}^v + v_k,\quad x_{k^*}^v=x_{k^*} \label{eq:state_noise}\\
&{x}_{k+1}^\delta = (F + GK)x_{k}^\delta - GK e_{k}^\delta,\qquad\quad\, x_{k^*}^\delta=\mathbf{0}. \label{eq:state_attack}
\end{align}

With these definitions, there are two reachable sets we aim to identify: the reachable states due to noise and due to attack. Because the state equation depends on the estimation error, in general, we must first identify the reachable estimation error due to noise and due to attack. Interestingly, the noise equations \eqref{eq:estimation_noise} and \eqref{eq:state_noise} have a special symmetry due to the zero initial conditions, i.e., $x_1^v=e_1^v=\mathbf{0}$. By writing out $e_k^v$ and $x_k^v$ for each $k=1,2,\dots$, it quickly becomes clear that $x_k^v=e_k^v$ for all $k\in\mathbb{N}$. Thus, for the contribution driven by noise, we need only to solve the $e_k^v$ equation. The reachable set of the estimation error driven by noise equals the reachable set of the states driven by noise.

Notice that the noise, a multivariate Gaussian distribution, also has unbounded support, thus it also has an infinite reachable set. We use the notion of a $\bar{p}$-probable reachable set to define a finite reachable set; for an equitable contribution by noise and attack, we again select $\bar{p}=1-\mathcal{A}$ and truncate the distribution with $\bar{v}$ such that
\begin{equation}
	\text{Pr}(v_k^TR_1^{-1}v_k \leq \bar{v}) = \bar{p} = 1-\mathcal{A},
\end{equation}
where $R_1$ is the covariance matrix of the system noise $v_k$. Since $v_k^TR_1^{-1}v_k$ is a chi-squared random variable with $n$ degrees of freedom, the value of $\bar{v}$ can be determined by Lemma \ref{lem:chisquared_tuning}.
Thus for noises, $k\in\mathbb{N}$,
\begin{align}
\mathcal{R}_x^{v} = \mathcal{R}_e^{v} &= \left\{ e_{k}^v \in \mathbb{R}^n \left| \ \eqref{eq:estimation_noise},\  v_k^TR_1^{-1}v_k \leq \bar{v} \right. \right\}. \label{eq:reachableset_estimation_noise}
\end{align}
For attacks, $\forall\ k \geq k^*$,
\begin{align}
\mathcal{R}_e^{\delta} &= \left\{ e_{k}^\delta \in \mathbb{R}^n \left| \ \eqref{eq:estimation_attack},\  \bar{\delta}_k^T \bar{\delta}_k \leq \alpha \right. \right\}, \label{eq:reachableset_estimation_attack}\\
\mathcal{R}_x^{\delta} &= \left\{ x_{k}^\delta \in \mathbb{R}^n \left| \ \eqref{eq:state_attack},\  e_k^\delta\in \mathcal{R}_e^\delta \right. \right\}. \label{eq:reachableset_state_attack}
\end{align}

\subsection{LMI Approach}
In general, it is analytically intractable to compute a reachable set $\mathcal{R}$ exactly. Instead, using Linear Matrix Inequalities (LMIs), for some positive definite matrix $\mathcal{P}$, we derive \emph{outer ellipsoidal bounds} of the form $\mathcal{E}=\{ \xi_{k} \ |\ \xi_{k}^T \mathcal{P} \xi_{k} \leq 1 \}$ containing $\mathcal{R}$. Our LMI results leverage the following lemma; this approach parallels work in \cite{Carlos_Justin3,RuthsACC_Hidden} however, the attack definitions are different, so they should be reformulated here.

\begin{lemma} \label{lem:that} \cite{Reachable_set_1}.
Let $V_k$ be a positive definite function, $V_1 = 0$, and $\zeta_k^T \zeta_k \leq \kappa \in \mathbb{R}_{>0}$. If there exists a constant $a \in (0,1)$ such that the condition below holds, then $V_k \leq 1$:
\begin{equation}\label{eq:that}
V_{k+1} - aV_k - \frac{1-a}{\kappa} \zeta_k^T \zeta_k \leq 0.
\end{equation}
\end{lemma}
\vspace{1ex}

We present a generic solution to identify the outer bounding ellipsoids we need. We consider a linear system driven by an input that is elliptically bounded, which, as we will show, represent the dynamics in \eqref{eq:estimation_noise}-\eqref{eq:state_attack} and the corresponding constraints in \eqref{eq:reachableset_estimation_noise}-\eqref{eq:reachableset_state_attack}.
\begin{proposition} \label{prop:generic_ellipsoid}
Given a LTI system $\xi_{k+1}=A\xi_k+B\mu_k$, $A\in\mathbb{R}^{n \times n}$ and $B\in\mathbb{R}^{n \times p}$, with the constraint $\mu_k^TR\mu_k\leq 1$, $R>0$, for all $k\in\mathbb{N}$, if there exists $a\in(0,1)$ and positive definite matrix $\mathcal{P}\in\mathbb{R}^{n\times n}$ that solves the convex optimization,
\begin{equation} \label{eq:convex_optimization}
\left\{\begin{aligned}
	&\min_{\mathcal{P}}\ -\log\det{\mathcal{P}},\\
    &\text{s.t.}\ \mathcal{P}>0,\ \text{and}\\
    &\quad \begin{bmatrix}
		a\mathcal{P} - A^T \mathcal{P} A & -A^T \mathcal{P} B \\ -B^T \mathcal{P} A & (1-a)R - B^T \mathcal{P} B
	 \end{bmatrix} \geq 0,
\end{aligned}\right.
\end{equation}
then the reachable states $\mathcal{R} \subseteq \mathcal{E} = \{ \xi_{k} \in \mathbb{R}^{n}\ |\ \xi_{k}^T \mathcal{P} \xi_{k} \leq 1 \}$ and the ellipsoid $\mathcal{E}$ has minimum volume.
\end{proposition}
\vspace{1ex}
\begin{proof}
Let $V_k=\xi_k^T\mathcal{P} \xi_k$ and $\zeta_k = R^{\frac{1}{2}}\mu_k$ in \eqref{eq:that} in Lemma \ref{lem:that}, where $R^{\frac{1}{2}}$ is the symmetric square root of the positive definite matrix $R$. It is easy to confirm that $\zeta_k^T\zeta_k = (R^{\frac{1}{2}}\mu)^T(R^{\frac{1}{2}}\mu) = \mu_k^TR\mu_k\leq 1$ with $\kappa=1$. Substituting the dynamic equation for $\xi_{k+1}$ in $V_{k+1}$ yields an expression that when factored into quadratic form $\nu_k^T Q \nu_k \geq 0$, with $\nu_k=[\xi_k^T, \mu_k^T]^T$, the matrix $Q$ is the LMI in the optimization problem above. Thus the bounding ellipsoid is given by $\mathcal{E}= \{\xi_k\ |\ V_k=\xi_k^T\mathcal{P}\xi_k \leq 1 \}$. 

To ensure that the ellipsoid bound is as tight as possible, we minimize $(\det{\mathcal{P}})^{-\frac{1}{2}}$ since this quantity is proportional to the volume of $\mathcal{E}$. We instead minimize $\log\det{\mathcal{P}^{-1}}$ as it shares the same minimizer and because for $\mathcal{P}>0$ this objective is convex \cite{BEFB:94}. 
\end{proof}
\vspace{1ex}

We now use this generic result to outer bound the four reachable sets we need ($I_n$ is the $n \times n$ identity matrix).
\vspace{1ex}

\begin{theorem}
The reachable sets 
$$\mathcal{R}_e^v=\mathcal{R}_x^v,\ \mathcal{R}_e^\delta,\ \text{and}\ \mathcal{R}_x^\delta,$$ are contained in the minimum volume ellipsoids 
$$\mathcal{E}_e^v=\mathcal{E}_x^v,\ \mathcal{E}_e^\delta,\ \text{and}\ \mathcal{E}_x^\delta,$$
respectively, characterized by the positive definite matrices 
$$\mathcal{P}_e^v=\mathcal{P}_x^v,\ \mathcal{P}_e^\delta,\ \text{and}\ \mathcal{P}_x^\delta,$$
respectively, which are the solutions to the convex optimization in Proposition \ref{prop:generic_ellipsoid} with to the following choices of $A$, $B$, and $R$, respectively:
\begin{itemize}\vspace{1ex}
\item $\mathcal{P}_e^v=\mathcal{P}_x^v$: $A=F$, $B=I_n$, $R = \frac{1}{\bar{v}}R_1^{-1}$, \vspace{1.5ex}
\item $\mathcal{P}_e^\delta$: $A=F$, $B=-L\Sigma^{\frac{1}{2}}$, $R = \frac{1}{\alpha}I_p$,\vspace{1.5ex}
\item $\mathcal{P}_x^\delta$: $A=F+GK$, $B=-GK$, $R=\mathcal{P}_e^\delta$.
\end{itemize}
\end{theorem}
\vspace{1ex}
\begin{proof}
The proofs of each case are quite similar. We prove the case for $\mathcal{P}_x^\delta$ and the rest follow a same pattern. In \eqref{eq:reachableset_state_attack}, we identified that $e_k^\delta\in\mathcal{R}_e^\delta$. Instead we impose $e_k^\delta\in\mathcal{E}_e^\delta$. These sets are not equal, but since the ellipsoid contains the reachable set, this still satisfies the requirement of \eqref{eq:reachableset_state_attack} - it does so with extra conservatism by also including estimation errors $e_k^\delta\in \mathcal{E}_e^\delta\setminus\mathcal{R}_e^\delta$.

Setting $A=F+GK$ is straightforward comparing \eqref{eq:state_attack} to Proposition \ref{prop:generic_ellipsoid}. Define the input vector $\mu_k=e_k^\delta$ such that
\begin{equation}
	\mu_k^TR\mu_k = \mu_k^T\mathcal{P}_e^\delta\mu_k \leq 1,
\end{equation}
since $(e_k^\delta)^T \mathcal{P}_e^\delta e_k^\delta \leq 1$ by the definition of the ellipsoid $\mathcal{E}_e^\delta$. With this definition of $\mu_k$, the corresponding input matrix that satisfies \eqref{eq:state_attack} is $B=-GK$.
\end{proof}
\vspace{1ex}

Having derived the set of reachable states due to noise and due to attack, both bounded by ellipsoids, we now compose these together to yield the total reachable set of states. The superposition of two ellipsoidal sets has been studied extensively and labeled the geometric (Minkowski) sum such that $\mathcal{E}_1\oplus\mathcal{E}_2 = \{ x+y\ |\ x\in\mathcal{E}_1,\ y\in\mathcal{E}_2 \}$. The complete reachable set is then $\mathcal{E}_x = \mathcal{E}_x^v \oplus \mathcal{E}_x^\delta$. It is possible to compose another convex optimization and LMI to combine the ellipsoids \cite{RuthsACC_Hidden}, the geometric sum provides a tighter resulting ellipsoid. When these techniques are used to design controller and estimator gains, optimization methods are preferred, but in this work we do not follow this line of inquiry.

\subsection{Geometric Approach}
\label{geosum}
A geometric approach to finding the ellipsoidal bounds for the reachable set of states comes from the observation that the equations in \eqref{eq:estimation_noise}-\eqref{eq:state_attack} contain inputs that are ellipsoidally bounded, i.e., $\frac{1}{\bar{v}}v_k^TR_1^{-1}v_k\leq 1$ and $\frac{1}{\alpha}\bar\delta_k^T\bar\delta_k \leq 1$ (in fact these are spherically bounded). The geometric sum introduced above provides an operation that simultaneously computes all possible combinations between two geometric sets. For example, the dynamics for $k\in\mathbb{N}$,
\begin{equation}
	\xi_{k+1} = A\xi_k + B\mu_k,\quad \xi_1 = \mathbf{0},\quad\text{with}\quad \mu_k^TR\mu_k \leq 1,
\end{equation}
can be interpreted as an ellipsoidal update. For $k=1$,
\begin{equation}
	\xi_2 = A\xi_1 + B\mu_1.
\end{equation}
The $\mu_1^TR\mu_1\leq 1$ bound identifies that any possible value of $\mu_1$ belongs to an ellipse $\mu_1\in\{\mu\ |\ \mu^TR\mu\leq 1\}$. Many ellipsoid calculations are more concise when the ellipsoid is characterized by its \textit{shape matrix}, $\mathcal{Q}$, $\mathcal{E}(\mathcal{Q})=\{\mu \ |\ \mu^T\mathcal{Q}^{-1}\mu\leq 1\}$. With this definition it is easy to express the linear transformation of an ellipse: if $\xi = M\mu$ and $\mu\in\mathcal{E}(\mathcal{Q})$, then $\xi\in\mathcal{E}(M\mathcal{Q}M^T)$ \cite{ellipsoidal_toolbox}. Thus in this example $\mu_1\in\mathcal{E}(R^{-1})$ and $\xi_2\in\mathcal{E}(BR^{-1}B^T)$. Continuing,
\begin{equation}
	\xi_3 = A\xi_2 + B\mu_2,
\end{equation}
where $\xi_2\in\mathcal{E}(BR^{-1}B^T)$ and $\mu_2^TR\mu_2\leq 1$. In this case $\xi_3\in \mathcal{E}(ABR^{-1}B^TA^T)\oplus\mathcal{E}(BR^{-1}B^T)$, where $\oplus$ represents the geometric sum. Although the geometric sum of two ellipsoids is not necessarily an ellipsoid, there are straightforward techniques to tightly fit an ellipsoid around the resulting shape (see details in \cite{ellipsoidal_toolbox}) so we will consider for the rest of this paper that the geometric sum of two ellipsoids produces an ellipsoid. It is worth noting that this fitting does embed an element of conservatism in the result due to the fitting; therefore, we will minimize the number of times the fitting needs to occur in our proposed algorithm. It is also important to note that the geometric sum of two ellipsoids is only valid if the two ellipsoids are independent. Here each ellipsoid corresponds to a different realization $\mu_k$.

We see now that all possible values that $\xi_k$ can take on will belong to an ellipsoid, and one that is iteratively updated along the lines of the discussion above. Now we will specialize these observations to the context here to find $\mathcal{E}_x$. We take the same approach as in the LMI method by splitting the dynamics into a contribution driven by noise and a contribution driven by the attack, such that again $\mathcal{E}_x = \mathcal{E}_x^v \oplus \mathcal{E}_x^\delta$.

\begin{theorem} \label{thm:estimation_geometric}
Given the estimation and state equations \eqref{eq:estimation_noise} and \eqref{eq:state_noise} for the system driven by noise, the ellipse $\mathcal{E}_x^v$ contains all possible values of $x_k^v$, where,
\begin{equation}\label{eq:noise_geometric}
	\mathcal{E}_x^v = \bigoplus_{k=0}^\infty \mathcal{E}\left(\bar{v} F^{k} R_1 (F^{k})^T\right).
\end{equation}
In other words $\mathcal{R}_x^v\subseteq\mathcal{E}_x^v$.
\end{theorem} 
\vspace{1ex}
\begin{proof}
Recall for the contribution driven by noise we need only to solve the $e_k^v$ equation \eqref{eq:estimation_noise}. Expanding the recursive definition we can express $e_N^v$ ($x_N^v$) in terms of all the past terms,
\begin{equation}
\label{eq:noise_recursive}
x_N^v=e_N^v=\sum_{k=1}^{N-1}F^{N-1-k}v_k.
\end{equation}
Note that $v_i$ and $v_j$ are independent and equivalently bounded. Therefore, the terms of \eqref{eq:noise_recursive} have identical ellipsoids, $\mathcal{E}(\bar{v}R_1)$ transformed by different powers of $F$,
\begin{align}
\mathcal{E}_{x_N}^v&=\bigoplus_{k=1}^{N-1} \mathcal{E}\left(F^{N-1-k}(\bar{v}R_1)(F^{N-1-k})^T \right), \nonumber\\
	&=\bigoplus_{k=0}^{N-2} \mathcal{E}\left(\bar{v}F^k R_1(F^k)^T \right). \label{eq:noise_recursive_geometric}
\end{align}
The ellipsoid that bounds all possible trajectories is the limiting ellipsoid as $N$ goes to infinity. 
\end{proof}
\vspace{1ex}

\begin{remark}
We assume the matrix $F$ is stable, otherwise the attacker can easily achieve arbitrarily large reachable sets simply by decoupling the controller from the open-loop system. Because of this, the volume of the ellipsoids (proportional to the determinant of their shape matrix) with higher powers of $F$ become vanishingly small,
\begin{equation}
	\det\left(\bar{v}F^kR_1(F^k)^T\right) = \bar{v}\det(R_1)(\det{F})^{2k}.
\end{equation}
Since $\rho[F]<1$, $\det{F}<1$ and the volume goes to zero as $k$ becomes large. The practical application of this is that one can simply take $N$ terms of the limiting geometric sum in \eqref{eq:noise_geometric} to achieve an accurate approximation of the bounding ellipsoid. The convergence of this sum (and hence the number of terms that should be chosen) depends on the spectrum of $F$.
\end{remark}
\vspace{1ex}

\begin{figure*}[t]
\centering
\includegraphics[width=0.9\linewidth]{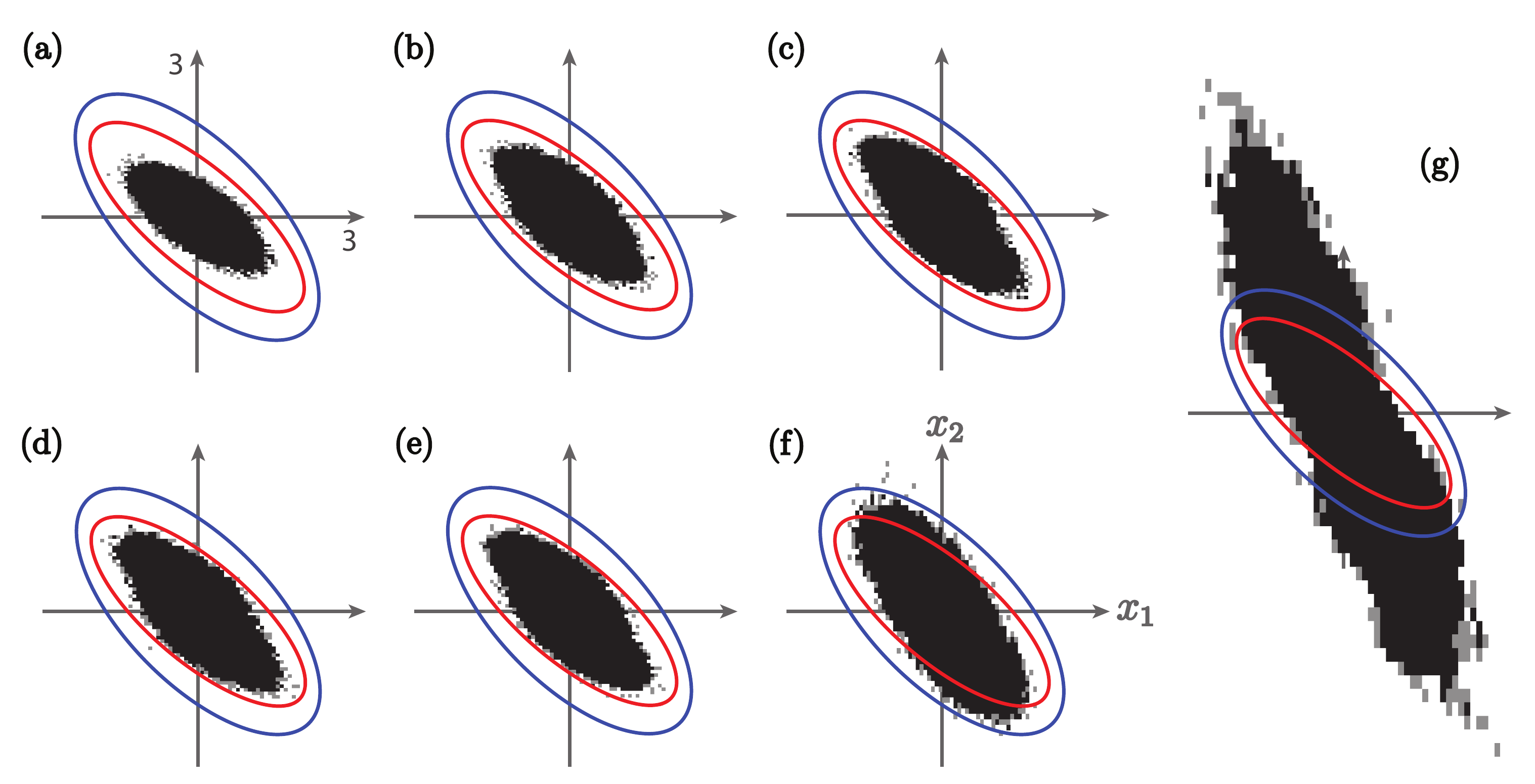}
\centering
\caption{The empirical reachable state sets in black with ellipsoidal bounds derived by the LMI approach (blue) and geometric approach (red) for zero alarm attacks (a) ZA.A, (b) ZA.B, and (c) ZA.C, as well as hidden attacks (d) H.A, (e) H.B, (f) H.C, and (g) H.D.} 
\label{fig:reachablesets}
\end{figure*}

\begin{figure}[t]
\centering
\includegraphics[width=\linewidth]{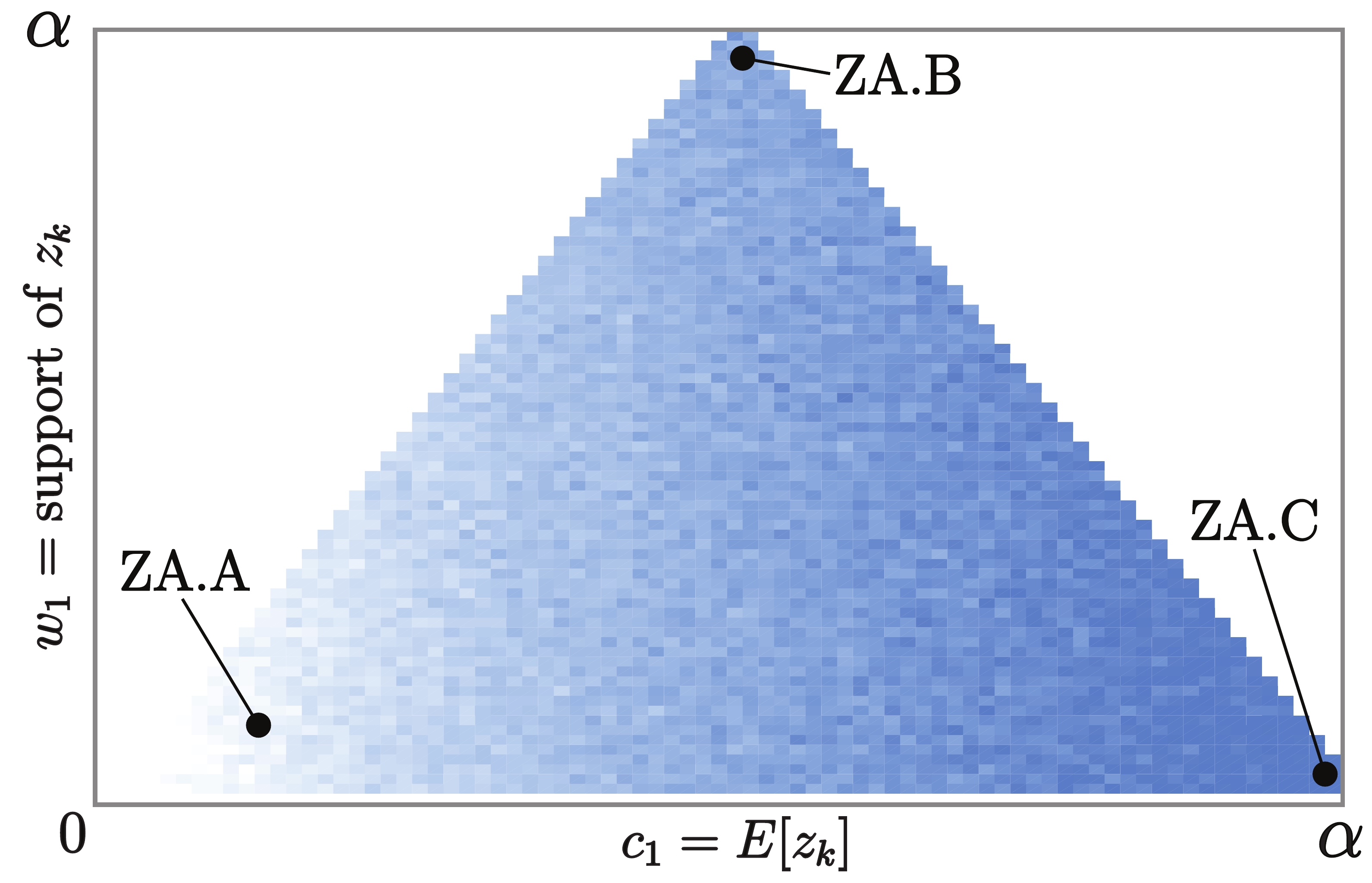}
\centering
\caption{The volume (blue is larger volume) of the reachable sets for different zero alarm $z_k$ distributions of the form shown in Table \ref{tbl:attack}.} \label{fig:triangle}
\end{figure}

\begin{theorem} \label{thm:attack_geometric}
Given the estimation and state equations \eqref{eq:estimation_attack} and \eqref{eq:state_attack} for the system driven by attack, the ellipse $\mathcal{E}_x^\delta$ contains all possible values of $x_k^\delta$, where,
\begin{equation}\label{eq:attack_geometric}
	\mathcal{E}_x^\delta = \bigoplus_{k=1}^\infty \mathcal{E}\left(\alpha H L\Sigma L^T H^T\right),
\end{equation}
where $H=(F+GK)^k-F^k$. In other words $\mathcal{R}_x^\delta \subseteq \mathcal{E}_x^\delta$.
\end{theorem} 
\vspace{1ex}

\begin{proof}
Expanding the recursive definitions in \eqref{eq:state_attack} and substituting in \eqref{eq:estimation_attack}, we find
\begin{equation} \label{eq:attack_recursive}
x_N^\delta=\sum_{k=1}^{N-2}\left((F+GK)^{N-1-k}-F^{N-1-k}\right)L\Sigma^{\frac{1}{2}}\delta_k.
\end{equation}
The rest of the proof follows the same line as the proof of Theorem \ref{thm:estimation_geometric} and so we omit the details. 
\end{proof}
\vspace{1ex}

\begin{remark}
Similarly, $\rho[F+GK]<1$ because the controller matrix is selected to make the closed-loop system stable. Thus Theorem \ref{thm:attack_geometric} benefits from the same practical advantage of constructing a good approximation of the ellipsoid $\mathcal{E}_x^\delta$ with finitely many terms.
\end{remark}

\section{Empirical Reachable Sets}
We now demonstrate these tools and provide a comparison between zero-alarm and hidden attacks. Although we generate a common bounding ellipsoid for both attacks, we also run extensive Monte-Carlo simulations to derive an approximation of the empirical reachable set the ellipsoids are meant to bound. We consider the following system for this study with the chi-squared detector tuned to a false alarm rate $\mathcal{A}=0.05$ (5\%):

{\footnotesize
\begin{align*}
&F= \begin{bmatrix}
       0.84 & 0.23 \\
       -0.47 & 0.12
     \end{bmatrix},\ 
G=\begin{bmatrix}
       0.07 & -0.32 \\
       0.23 &  0.58
     \end{bmatrix},\ 
C= \begin{bmatrix}
       1 & 0 \\
       2 &  1
     \end{bmatrix}, \\ \vspace{3ex}
&R_1= \begin{bmatrix}
       0.045 & -0.011\\
       -0.011 & 0.02
     \end{bmatrix}, \ 
K= \begin{bmatrix}
       1.404 & -1.042 \\
       1.842 &  1.008
     \end{bmatrix},\\ \vspace{3ex}
&L= \begin{bmatrix}
       0.0276 & 0.0448 \\
       -0.01998& -0.0290
     \end{bmatrix},\ 
R_2=\begin{bmatrix}
       2 & 0\\
       0 & 2
     \end{bmatrix}, \ 
\Sigma= \begin{bmatrix}
       2.086 & 0.134 \\
       0.134 &  2.230
     \end{bmatrix}.
\end{align*}}

\begin{table}[t]
\centering
\caption{Parameters for zero alarm (ZA) and hidden (H) attacks.}
\vspace{-3mm}
\includegraphics[width=0.85\linewidth]{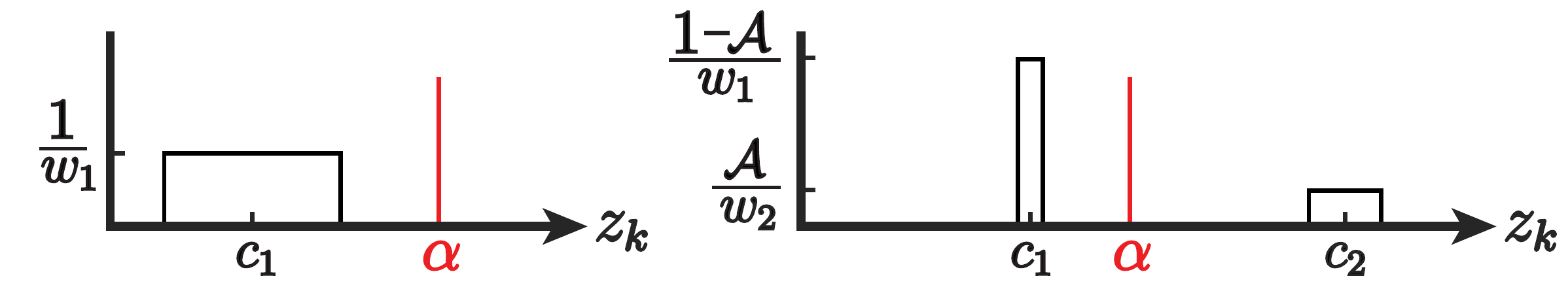}\\
\vspace{1mm}
\begin{tabular}{r|c|c|c|c|c|c|c}
 & ZA.A & ZA.B & ZA.C & H.A & H.B & H.C & H.D \\ \hline
$c_1$ &  ${\alpha}/{8}$ & ${\alpha}/{2}$ & $\alpha$ & $\alpha$ & $\alpha$ & $\alpha$ & $\alpha$ \\
$w_1$ &  ${\alpha}/{10}$ & $\alpha$ & 0 & 0 & 0 & 0 & 0 \\
$c_2$ &-&-&-& $1.5\alpha$ & $2\alpha$ & $10\alpha$ & $100\alpha$ \\
$w_2$ &-&-&-& $\alpha$ & 0 & 0 & 0
\end{tabular} \label{tbl:attack}
\end{table}

Fig. \ref{fig:attack_distributions} clearly shows the ambiguity in designing zero alarm attacks and hidden attacks.  In a zero alarm attack, the density function can be arbitrarily shaped on $z_k\leq\alpha$. For simplicity, consider that we use a uniform distribution of width $w_1$ and centered at $z_k=c_1$, such that the support of the distribution is over $[c_1-\frac{w_1}{2},c_1+\frac{w_1}{2}]$. In other work, we have shown that in terms of steady-state deviation of the state, there exists a magnitude and ``direction'' of $\bar\delta_k$ that yields the strongest attack \cite{RuthsACC_Windowed}. It is intuitive that maximizing the norm of the attack $\bar\delta_k^T\bar\delta_k=\alpha$ leads to stronger attacks than $\bar\delta_k^T\bar\delta_k<\alpha$. While we show an analytic result for the steady state deviation of the state \cite{RuthsACC_Windowed}, showing the same for the reachable sets is more nuanced. To demonstrate that this intuition holds empirically, we calculate the volume of the empirical reachable set attained through simulation. We approximate the volume by fitting an ellipsoid to the point cloud, where the lengths of the principle axes are given in terms of the eigenvalues of the data, as $1/\sqrt{\lambda_1}$ and $1/\sqrt{\lambda_2}$. The volumes are plotted in Fig. \ref{fig:triangle} for different choices of $w_1$ and $c_1$ and clearly shows the largest volume ellipsoids (dark blue) are generated by attacks for which $c_1=\alpha$ and $w_1\approx 0$. We select three sets of values for the pair $(w_1,c_1)$ labeled ZA.A, ZA.B, and ZA.C for our comparison (see Table \ref{tbl:attack}).

For hidden attacks, there are two regions to define (below and beyond $z_k=\alpha$). Based on our observations (and intuition), we set the $1-\mathcal{A}$ portion of the distribution that falls at or below $\alpha$ as a point mass at $\alpha$. From Fig. \ref{fig:triangle}, an attacker who wishes to maximize their influence on the reachable set would naturally make this choice. As discussed before, the second mass lies beyond $\alpha$ and could theoretically cause arbitrarily large reachable sets. Here we select the $\mathcal{A}$ mass in four different configurations (parameterized by a second uniform distribution section centered at $c_2$ and with width $w_2$, see Table \ref{tbl:attack}): spread uniformly from $(\alpha,2\alpha]$ (labeled H.A), a point mass at $2\alpha$ (H.B), a point mass at $10\alpha$ (H.C), and a point mass at $100\alpha$ (H.D).

In Fig. \ref{fig:reachablesets}, we display the empirical reachable sets for all seven of these attacks as well as the LMI (blue) and geometric (red) outer ellipsoidal bounds derived with our methods. We first observe that both techniques are able to rather tightly bound the reachable set of states due to zero alarm attacks, although the geometric approach provides slightly tighter ellipsoid bounds. For hidden attacks, while it takes high magnitude attacked $z_k$ values (e.g., $c_2=10\alpha,\ 100\alpha$) to see a distinct growth in the volume of the reachable set, it is possible to grow the reachable set arbitrarily large.

The more substantial takeaway from this empirical study is that when we use conventional detectors that use a single cut in the distribution to determine if the current $z_k$ is more likely to come from the original attack-free distribution or some other (attacked) distribution, we lack the ability to constrain the attacker due to the $\mathcal{A}$ fraction of the distribution that falls beyond the detector threshold $\alpha$. We require either a combination of detectors or modified definitions of current detectors to synthesize the information necessary to limit attackers further. When attackers hide in the infinite support of the noise, as in a hidden attack, we require some mechanism to effectively truncate or bound the impact of an attacker. Some obvious solutions are available, such as enforcing finite support of all noises, however, these approaches have not been integrated into conventional detector methods. In addition saying a disturbance as finite support is different from practically using this assumption; this gap must be addressed before this type of approach could be used.

\section{CONCLUSION}
We have presented a thorough exposition of the current ideology on using fault detection type detectors for identification of (sensor) attacks on control systems. In particular, we compared two attacks in which the opponent aims to remain stealthy - one in which the attack sequence is generated so as to not raise any alarms and one in which the attack sequence raises alarms at the same rate they occur randomly in the absence of attacks. We developed two approaches to determine ellipsoidal and $\bar{p}$-probable ellipsoidal bounds (when the reachable set is infinite) on the reachable states of the system in response to the attack and to the inherent system noise. We demonstrated these concepts and methods with a numerical example that emphasizes the need for work that goes beyond traditional detectors.



\bibliographystyle{IEEEtran}
\bibliography{security}
\end{document}